\documentclass[conference, 10pt]{IEEEtran}
\usepackage{cite}
\usepackage{url}
\usepackage[absolute]{textpos}
\newcommand{\copyrightstatement}{
    \begin{textblock}{15}(0.5,0.3)    
         \noindent
         \centering
         \textblockcolour{white}
         \footnotesize
         \copyright 2016 IEEE. Personal use of this material is permitted. Permission from IEEE must be obtained for all other uses, in any current or future media, including reprinting/republishing this material for advertising or promotional purposes, creating new collective works, for resale or redistribution to servers or lists, or reuse of any copyrighted component of this work in other works
    \end{textblock}
}

\ifCLASSINFOpdf
\else
 \usepackage[dvips]{graphicx}
  \DeclareGraphicsExtensions{.eps}
\fi

\usepackage{psfrag}

\usepackage[cmex10]{amsmath}

\usepackage{graphicx}
\usepackage{color}
\usepackage{placeins}
\usepackage{float}
\usepackage{tabularx,colortbl}

\begin{document}
%
\title{A Bitstream Feature Based Model for Video Decoding Energy Estimation}

\copyrightstatement

\author{\IEEEauthorblockN{Christian Herglotz, Yongjun Wen, Bowen Dai, Matthias Kr\"anzler, Andr\'e Kaup}
\IEEEauthorblockA{Chair of Multimedia Communications and Signal Processing\\
Friedrich-Alexander University Erlangen-N\"urnberg (FAU)\\
Cauerstr. 7, Erlangen, Germany\\
\{ christian.herglotz@, yongjun.wen@studium., bowen.dai@studium, matthias.kraenzler@, andre.kaup@ \}  FAU.de\\}}



\renewcommand{\baselinestretch}{0.95}

\maketitle

\begin{abstract}
In this paper we show that a small amount of bit stream features can be used to accurately estimate the energy consumption of state-of-the-art software and hardware accelerated decoder implementations for four different video codecs. By testing the estimation performance on HEVC, H.264, H.263, and VP9 we show that the proposed model can be used for any hybrid video codec. We test our approach on a high amount of different test sequences to prove the general validity. We show that less than $20$ features are sufficient to obtain mean estimation errors that are smaller than $8\%$. Finally, an example will show the performance trade-offs in terms of rate, distortion, and decoding energy for all tested codecs. 
\end{abstract}


\IEEEpeerreviewmaketitle

\section{Introduction}
\label{sec:intro}
Recent studies showed that the global demand for video streaming applications is growing rapidly \cite{cisco16}. Video data already constitutes more than $50\%$ of the total internet traffic, where a major part of this traffic is used for mobile streaming applications on portable devices like smartphones or tablet PCs. Unfortunately, these devices are battery driven such that the available processing power is limited. Hence, research aiming at reducing the energy consumption of the video streaming process is a worthwhile task. 

To this end, researchers developed different methods to reduce the energy consumption of video decoders. The most popular method is to create dedicated decoding hardware. E.g., Engelhardt et al. developed an HEVC-decoder for FPGA \cite{Engelhardt14}. In other works, dedicated hardware modules for major decoder functions such as the deblocking filter \cite{Adibelli11} or the integer transform \cite{Do14} 
are presented. In a different direction, the processing energy is reduced using dynamic voltage and frequency scaling to reduce the power consumption of the CPU \cite{Akyol07}. Finally, research has been performed on the complexity of the decoding process where the encoder adopts the estimated decoding complexity into the rate-distortion optimization formula \cite{Lee07b}. Similarly, a dedicated model for estimating the energy consumption of an HEVC decoder \cite{Herglotz14} was developed and successfully applied in the encoder to produce bit streams that require less decoding energy at the same objective visual quality \cite{Herglotz16a}. 

In this paper, we generalize the model for estimating the HEVC decoding energy presented in \cite{Herglotz14} to be applicable to other codecs and decoder implementations. The model is based on bit stream features that describe the main processing steps that are executed repeatedly during the decoding process. For a given input bit stream, the decoding energy is estimated by 
\begin{equation}
\hat E = \sum_{f=1}^F n_{f} \cdot e_{f},  
\label{eq:generalModelFunction}
\end{equation}
where $f$ is the feature index, $F$ the magnitude of the feature set that depends on the used codec, $n_{f}$ the feature number that describes how often feature $f$ occurs, and $e_{f}$ the feature's specific energy that represents the mean processing energy consumed upon each occurence of the corresponding feature. By generalizing this model to other codecs, in this paper we will 
\begin{itemize}
\item show the general applicability of the model, 
\item provide a new method to compare the energetic properties of different video codecs, 
\item construct models that can be further used for decoding energy optimization. 
\end{itemize}
Particularly, we will adapt this model to the H.263 \cite{ITU_H.263}, the H.264 \cite{ITU_H.264}, and the VP9 codec \cite{Mukherjee13} and, by comparing to the baseline HEVC model, show that the proposed feature based model can be used to accurately estimate the decoder's energy consumption independent from the used codec and its implementation. 

The paper is organized as follows: Section \ref{sec:features} gives more details on the general concept of a bit stream feature and introduces the different feature categories that are used. Then, Section \ref{sec:sets} presents the explicit feature sets used for the codecs and explains the corresponding feature analyzers used to determine the feature numbers $n_f$. Afterwards, Section \ref{sec:eval} introduces our evaluation setup and gives a thorough analysis on the test results. Section \ref{sec:concl} concludes the paper. 

\section{Bit Stream Features}
\label{sec:features}
Generally, a bit stream feature describes the execution of a standardized process when decoding a given bit stream. As an example, one feature corresponds to the execution of a DCT-transform of a certain block size ({feature \footnotesize{trans}}). This transform is executed repeatedly during the decoding process and, due to the predefined processing flow, requires roughly the same amount of processing energy upon each execution. Hence, counting the number of executed transforms and determining the mean processing energy we can estimate the complete decoding energy related to transformation. 

Likewise, further features are defined where the feature numbers can be determined for any given, standard compliant bit stream. 
As all considered codecs use a block-based hybrid approach, we find that a general categorization of features can be defined: 
\begin{itemize}
\item \textbf{Offset features} (OFFSET): In this category, two features are defined that comprise processes during encoding that cannot be skipped. The first feature ({\footnotesize{$f=E_0$}}) 
corresponds to the offset energy required for starting and ending the decoding process. Hence, for each bit stream, the corresponding feature number is fixed to one. Secondly, the number of frames is counted to represent the processing energy used to initialize a frame ({\footnotesize{$f=$ frame}}). 
\item \textbf{Intraframe prediction features} (INTRA): These features correspond to all processes performed for intraframe prediction of a block ({\footnotesize{intra}}). 
Our studies showed that processing larger blocks requires less energy than using small blocks, hence we consider the different intraframe-prediction block sizes that the codec allows. 
\item \textbf{Interframe prediction features} (INTER): Similar to the intraframe case, these features describe the interframe prediction process of a certain block size ({\footnotesize{inter}}). 
To represent the motion compensation process, we count the number of pels that need to be predicted ({\footnotesize{pel}}), 
that are counted twice in biprediction. As additionally all codecs allow fractional pel motion vectors for more accurate motion compensation, the number of fractional pel filterings is counted ({\footnotesize{frac}}), too. 
 \item \textbf{Residual transformation features} (TRANS): For these features we count the number of inverse transforms performed for reconstruction ({\footnotesize{trans}}). 
Just like for prediction, we consider the transformation block size. 
\item \textbf{Residual coding features} (COEFF): To represent the parsing process of the residual coefficients we count the number of non-zero coefficients ({\footnotesize{coeff}}) 
and consider their value ({\footnotesize{val}}). 
\end{itemize}
Naturally, decoding consists of many more processes like loop filtering, de-quantization, motion vector coding and so forth. These processes can either be assigned to one of the above-mentioned features (like de-quantization to coefficient decoding as de-quantization must be performed for each non-zero coefficient), or they only consume a marginal amount of energy such that their consideration would not increase the estimation accuracy significantly. The explicit feature sets for the codecs are defined in the next section.

\section{Feature Sets}
\label{sec:sets}
In this section, the properties of the feature sets for each codec are discussed in detail. 
The most important difference results from the block sizes that are allowed. For intraframe prediction, all considered codecs only allow square blocks. H.263 provides a single block size ($16\times 16$), H.264 additionally uses $4\times 4$-blocks (we do not consider the high profile), and HEVC as well as VP9 allow four different sizes ($32\times 32$ to 
$4\times 4$). Note that in our work, we count the block size that is actually processed. E.g., in the HEVC standard a prediction block is allowed to be of size $64\times 64$, nevertheless processing is performed on four $32\times 32$ blocks corresponding to the residual transformation process 
\cite{Sullivan12}. 

For interframe prediction, even more block sizes are allowed. Including rectangular splits, e.g., VP9 includes $13$ different block sizes to choose from. To prune the resulting high amount of features, we propose merging some of the block sizes having a similar amount of pixels. As a result, we define two block sizes for H.263 ($16\times 16$ and $8\times 8$) and three for H.264 ($16\times 16$ to 
$4\times 4$), where the rectangular sizes are added with a half-weight to the next bigger square block. E.g., a block of size $8\times 16$ is counted as a half $16\times 16$ block. For HEVC, four block sizes are counted ($64\times 64$ to 
$8\times 8$) and for VP9, we additionally count $4\times 4$-blocks. Note that we define a special block for the H.263 standard that corresponds to overlapped block motion compensation ({\footnotesize{OBMC}}), cf. \cite{ITU_H.263}, as the corresponding process is more complex than standard motion compensation. 

For transformations, H.263 only provides a single block size ($8\times 8$). In H.264, only $4\times 4$ transformations are counted as we do not consider the high profile. For HEVC and VP9, the four transform sizes $32\times 32$ to 
$4\times 4$ are taken into account. 
\begin{table}[t]
\renewcommand{\arraystretch}{1.3}
\caption{Categorized feature sets for each codec. The numbers indicate how many block sizes are considered, if applicable. The magnitude of the feature sets is written in the last row. }
\vspace{-.5cm}
\label{tab:featList}
\begin{center}
\begin{tabular}{l|l|c|c|c|c}
\hline
Category & Feature $f$ & H.263 & H.264 & HEVC & VP9 \\
 \hline\hline
OFFSET&{\footnotesize{$E_0$}} & $1$&$1$&$1$&$1$\\
&{\footnotesize{frame}} &$1$&$1$&$1$&$1$\\
\hline
INTRA&{\footnotesize{intra}} & $1$ & $2$ & $4$ & $4$\\
\hline
&{\footnotesize{inter}} & $2$ & $3$ & $4$ & $5$\\
INTER&{\footnotesize{OBMC}} & $1$ & - &-&-\\
&{\footnotesize{pel}} & $1$&$1$&$1$&$1$\\
&{\footnotesize{frac}} &$1$&$1$&$1$&$1$\\
\hline
TRANS&{\footnotesize{trans} }& $1$ & $1$ & $4$ & $4$\\
\hline
COEFF&{\footnotesize{coeff}} & $1$ & $2$ & $1$ & $1$\\
&\footnotesize{val} & $1$ & $2$ & $1$ & $1$\\
\hline
SAO&{\footnotesize{SAO}} & - & - & $1$ & -\\
\hline\hline
&magn($F$)
& $11$ & $14$ & $19$ & $19$\\
\hline
\end{tabular}
\vspace{-.5cm}
\end{center}
\end{table}
In order to consider the value of the residual coefficients ({\footnotesize{val}}), 
two different methods are applied. For HEVC, we sum up the logarithms to the basis $2$ of each non-zero residual coefficient (cf. \cite{Herglotz16a}). For the other codecs, the number of coded bits for each coefficient is used. 

Finally, we would like to mention two special cases. The first case holds for H.264 where two different arithmetic coding methods (CAVLC and CABAC) are allowed. Therefore, the two residual coding features {\footnotesize{coeff}} and {\footnotesize{val}}  
are defined twice, once for CAVLC and once for CABAC. For the second case, we take the sample adaptive offset filter (SAO) \cite{Fu12} into account that was newly introduced for the HEVC standard. As SAO introduces a significant amount of additional complexity into the decoder, we count the number of $64\times 64$-sized luma blocks that are filtered by this new algorithm. 

Summarizing, all the above mentioned feature sets including their categorization are listed in Table \ref{tab:featList}. We can see that the recent codecs have a larger feature set which is caused by a higher amount of coding modes. The feature analyzers used to count the feature numbers are implemented into readily available, existing decoder solutions which are the TMN-2.0 \cite{TMN-2.0}, JM-18.4 \cite{JM}, HM-11.0 \cite{HM}, and libvpx \cite{libvpx} software decoders. In the next section we will show that these feature sets suffice to accurately estimate the decoding energy consumption.

\section{Evaluation}
\label{sec:eval}
In this section we thoroughly explain our evaluation method. First, the setup for measuring the decoding energy is introduced followed by a discussion on the evaluation sequences. Afterwards, the training and validation method is described and the final results are given. 

\subsection{Measurement Setup}
To measure the true decoding energies, the test setup presented in \cite{Herglotz16a} is used. As a power meter, we employ ZES Zimmer's LMG95 to obtain the energies $E_\mathrm{dec}$ required to decode all tested bit streams.
We measure the energy consumption of the FFmpeg software decoder \cite{FFmpeg} which is readily capable of decoding all considered codecs. We consider FFmpeg to show a realistic processing flow as it is optimized for real-time, practical applications. The decoding device (DEC) is a Pandaboard \cite{Panda} which features a smartphone like architecture using an ARM processor. 

To show that our model is not restricted to a single software solution, we evaluate the estimation accuracy on alternative decoders, namely the TMN-2.0 decoder for H.263 and libde265 \cite{libde} for HEVC. For H.264, a hardware accelerated decoder was tested on the Pandaboard which is included on the OMAP4430 system-on-chip (SoC) \cite{OMAP4430_tech}. It includes an image and video acceleration unit (IVA) that is capable of decoding H.264-coded videos with a resolution up to 1080p. 

\subsection{Evaluation Sequence Set}
The input sequences for H.264, HEVC, and VP9 are taken from all classes of the standard HEVC test set \cite{Bossen13} and are listed in Table \ref{tab:HEVCseq}. Furthermore, as H.263 only allows specific resolutions, further sequences are coded for all four codecs in CIF and QCIF resolution (Table \ref{tab:cifSeq}). 
\begin{table}[t]
\renewcommand{\arraystretch}{1.3}
\caption{Evaluation sequences taken from the HEVC test set. Except for Class A ($8$ frames), all sequences were coded using $40$ frames. }
\vspace{-.5cm}
\label{tab:HEVCseq}
\begin{center}
\begin{tabular}{c|c}
\hline
Class A ($2560\times 1600$ pixels) & Class B ($1920\times 1080$ pixels) \\
 \hline
PeopleOnStreet & BasketballDrive \\
Traffic & BQTerrace \\
 & Cactus \\
  & Kimono \\
   & ParkScene\\
   \hline\hline
   Class C ($832\times 480$ pixels) & Class D ($416\times 240$ pixels) \\
   \hline
   BasketballDrill & BasketballPass \\
   BQMall & BlowingBubbles \\
   PartyScene & BQSquare \\
   RaceHorses & \\
   \hline\hline
   Class E ($1280\times 720$ pixels) & Class F (variable resolution) \\
   \hline
   FourPeople & SlideEditing\\
   Johnny & SlideShow \\
   KristenAndSara & ChinaSpeed \\
   vidyo1,3,4 & BaketballDrillText \\
   \hline

\end{tabular}
\end{center}
\vspace{-.3cm}
\end{table}
\begin{table}[t]
\renewcommand{\arraystretch}{1.3}
\caption{Evaluation sequences compatible with H.263. 
}
\vspace{-.5cm}
\label{tab:cifSeq}
\begin{center}
\begin{tabular}{c|c}
\hline
QCIF ($176\times 144$ pixels) & CIF ($352\times 288$ pixels) \\
 \hline
Akiyo ($30$ frames)	& Foreman ($30$ frames)\\
Crew ($50$ frames) & Tennis ($30$ frames) \\
Miss America ($50$ frames) & Car Phone ($50$ frames) \\
Coastguard ($50$ frames) & Bus ($50$ frames) \\
News ($30$ frames) & Suzie ($30$ frames) \\
\hline
\end{tabular}
\end{center}
\vspace{-.4cm}
\end{table}

To take different visual qualities into account, for each codec a set of QPs was chosen that spans the range of PSNRs from around $25$dB to $55$dB. 
For encoding, the reference software encoders were chosen using different standard encoder configurations. Note that class A sequences were not tested for the H.264 hardware accelerated decoder as the resolution is too high. Detailed information about the coded sequences for all codecs is summarized in Table \ref{tab:config}. 

\begin{table}[t]
\renewcommand{\arraystretch}{1.3}
\caption{Software and configurations for encoding the evaluation bit streams. The last row in each column denotes the total number of tested bit streams. To obtain a sufficiently large test set for H.263, different parts of the input sequences were coded. }
\label{tab:config}
\vspace{-.5cm}
\begin{center}
\begin{tabular}{l|c|c}
\hline
 & H.263 & VP9  \\
 \hline
 Encoder & TMN-2.0 \cite{TMN-2.0} & libvpx \cite{libvpx} \\
 Configurations & Frames $1$ to $N$ & One-pass coding \\
  & Frames $N+1$ to $2N$ & Two-pass coding \\
  QPs & $1$, $3$, $7$, $12$, $23$, $30$ & $5$, $20$, $44$, $59$\\
  \# Bit streams & $120$ & $272$ \\
  \hline \hline
  & H.264 & HEVC \\
  \hline
  Encoder & JM-18.4 \cite{JM} & HM-16.4 \cite{HM} \\
  Configurations & baseline & intra \\
  & main & lowdelay \\
  & extended & lowdelay\_P \\
  & & randomaccess \\
  QPs & $12$, $22$, $32$, $42$ & $10$, $20$, $30$, $40$\\
  \# Bit streams & $408$ & $544$ \\
  \hline
\end{tabular}
\end{center}
\end{table}

\subsection{Validation Method}
\begin{figure}[t]
\centering
\psfrag{A}[c][b]{Coded evaluation sequences (Tables \ref{tab:HEVCseq}, \ref{tab:cifSeq})}
\psfrag{B}[c][c]{Bit stream analysis}
\psfrag{C}[c][c]{Measurement}
\psfrag{D}[c][c]{Bit stream specific variables $n_f$}
\psfrag{E}[c][c]{Decoding energies $E_\mathrm{dec}$}
\psfrag{F}[c][c]{Training}
\psfrag{G}[c][c]{Validation}
\psfrag{H}[c][c]{Model}
\psfrag{I}[c][c]{parameters $e_f$}
\psfrag{J}[l][l]{$10$-fold cross-validation}
\psfrag{K}[c][c]{Estimation error $\overline{\varepsilon}$}
\includegraphics[width=0.47\textwidth]{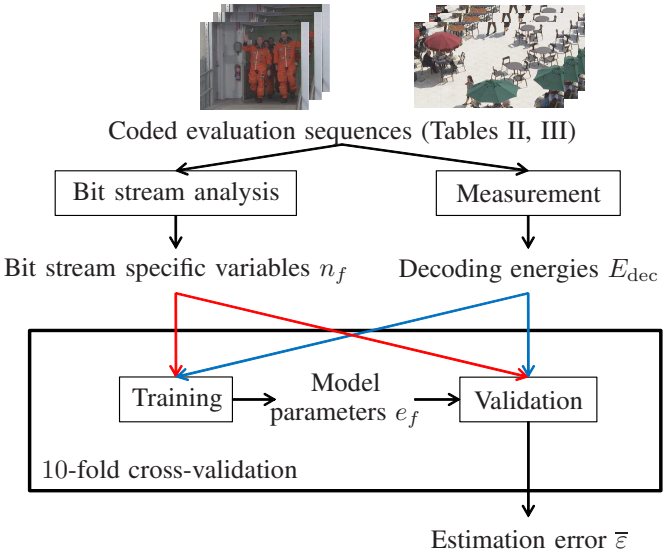}
\vspace{-.5cm}
\caption{Evaluation flow. The evaluation sequences are analyzed for their corresponding feature numbers $n_f$. Furthermore, the decoding energy of these sequences is measured for each codec using the setup shown in \cite{Herglotz16b}. Afterwards, we feed the bit stream specific variables and the decoding energies into a 10-fold cross-validation loop to train the model parameters and validate the estimation accuracy. As an output, we obtain the mean absolute estimation error $\overline{\varepsilon}$.  }
\label{fig:eval_flow}
\vspace{-.5cm}
\end{figure}

Our method for testing the estimation performance of the proposed model is depicted in Figure \ref{fig:eval_flow}. For each codec we perform a $10$-fold cross-validation as proposed in \cite{Zaki14}. In this method, we randomly divide the complete set of bit streams into $10$ approximately equally sized subsets. Then, we perform $10$ iterations where in each iteration, one subset is defined as the validation subset. The remaining $9$ subsets are used to train the parameters (which are the specific energies $e_f$ in our case) using the measured decoding energies $E_\mathrm{dec}$. As a training method we use the trust-region approach \cite{Coleman96} which aims at minimizing the squared error. The resulting specific energies $e_f$ are then used for validating the validation subset. 

As a criterion to express the estimation performance we use the mean relative estimation error calculated by 
\begin{equation}
\bar \varepsilon = \frac{1}{M}\sum_m{ \frac{\left|\hat E_m - E_{m,\mathrm{dec}}\right|}{E_{m,\mathrm{dec}}}},
\end{equation}
where $m$ is the bit stream index, $M$ the magnitude of the bit stream sets, and $\hat E_m$ and $E_{m,\mathrm{dec}}$ the estimated and measured decoding energy of the $m$-th bit stream, respectively. A mean relative error of $\bar \varepsilon = 0$ would indicate that we have a perfect estimator. 

To further prove the superior estimation performance of the proposed feature based model, we compare the results to the estimation error of two models from the literature. The first (HL1) is proposed in \cite{Herglotz15c} and estimates the decoding energy based on high-level properties of a coded bit stream which are resolution $S$, number of frames $N$, and bit stream file size $B$. In this model, the energy is estimated by 
\begin{equation}
\hat E_\mathrm{HL1} = C + S\cdot N\cdot \left[ \alpha + \beta \cdot \left(\frac{B}{S\cdot N}\right)^\gamma \right], 
\label{eq:HL_model}
\end{equation}
where the parameter $C$ can be interpreted as a constant offset energy, $\alpha$ as the offset energy needed to decode a pixel, and $\beta$ and $\gamma$ represent a pixel-wise additive term that depends on the mean amount of bits that are used for coding a pixel. 

The second model (HL2) was introduced in \cite{Raoufi13} and further refined in \cite{Herglotz16b} and reads 
\begin{equation}
\hat E_\mathrm{HL2} = \left(c_1\cdot p_\mathrm{I} \cdot \frac{B}{S\cdot N} + c_2 \cdot p_\mathrm{I} + c_3 \cdot \frac{B}{S\cdot N} + c_4 \right) \cdot N \cdot S. 
\label{eq:Raoufi}
\end{equation}
In addition to bitrate $B$, number of frames $N$, and resolution $S$ the rate of intra frames $p_\mathrm{I}$ (which is the number of intra frames divided by the complete number of frames) is considered. The parameters $c_1$ to $c_4$ are the codec and implementation specific parameters. Similar to the evaluation of the feature based model, the parameters of these two models are determined using the 10-fold cross-validation approach. 

\subsection{Results}
The resulting estimation errors are summarized in Table \ref{tab:errors}. 
\begin{table}[t]
\renewcommand{\arraystretch}{1.3}
\caption{Mean relative estimation errors $\bar \varepsilon$ for the four considered codecs. The first row gives the results for the proposed model. The second row validates the estimation errors on other decoder implementations. The last two rows show the estimation errors of the reference models.  }
\label{tab:errors}
\vspace{-.6cm}
\begin{center}
\begin{tabular}{l|l|c|c|c|c}
\hline
Model & Software & HEVC & H.264 & H.263 & VP9 \\
 \hline
Feature based & FFmpeg & $5.27\%$ & $6.41\%$ & $2.51\%$ & $5.11\%$\\ 
 Feature based & Alt.   & $3.18\%$ & $7.50\%$ & $0.77\%$  & - \\
 HL1    & FFmpeg & $13.04\%$ & $15.22\%$ & $2.67\%$ & $30.55\%$\\  
 HL2 & FFmpeg & $20.86\%$ & $26.23\%$ & $30.38\%$ & $33.34\%$\\  
 \hline
\end{tabular}
\vspace{-.5cm}
\end{center}
\end{table}
Considering the first row we can see that the proposed model reaches errors that are smaller than $7\%$ for all codecs. The lowest error is obtained for the H.263 codec ($2.51\%$) which is not surprising as it is the oldest one providing least complexity. Furthermore, estimation errors are significantly lower than errors returned by the high-level models (rows three and four) which can be explained by the increased number of parameters. Note that for HL2, under certain circumstances estimation errors of around $20\%$ can be reached. For H.264, the estimation error is smaller than $20\%$ when the bit streams are solely coded with the main encoder configuration and for H.263, the error is smaller than $8\%$ when only a single resolution is considered. Furthermore, the estimation errors of HL1 and HL2 for VP9 are relatively high ($>30\%$) which can be explained by the following two reasons: 
\begin{itemize}
\item Compound prediction: VP9 offers the possibility to code frames that are only used for prediction and not displayed. These additionally coded frames are not considered in the reference model. 
\item In \cite{Herglotz15c} it is stated that screen content videos are badly estimated by a high-level model. Removing them from the evaluation test set we could achieve estimation errors smaller than $22\%$ for both HL1 and HL2. 
\end{itemize}
In contrast, note that the estimations for the H.263 decoder given by HL1 are very sound ($2.64\%$) which can again be explained by the low complexity of this codec. 

The second row shows the estimation errors for the alternative software decoders. We can see that the values for H.263 and HEVC are even smaller ($3.18\%$ and $0.77\%$, respectively) which indicates that the general purpose FFmpeg solution is more difficult to model. As additionally the hardware accelerated decoder shows a relatively low estimation error ($7.5\%$), we can say that the proposed model can be used independent from the decoder implementation. 

\begin{figure}[t]
\centering
\psfrag{000}[c][c]{$0$}
\psfrag{001}[c][c]{$0.2$}
\psfrag{002}[c][c]{$0.4$}
\psfrag{003}[c][c]{$0.6$}
\psfrag{004}[c][c]{$0.8$}
\psfrag{022}[l][l]{$E_\mathrm{dec}$}
\psfrag{021}[l][l]{OFFSET}
\psfrag{020}[l][l]{INTRA}
\psfrag{019}[l][l]{INTER}
\psfrag{018}[l][l]{TRANS}
\psfrag{017}[l][l]{COEFF}
\psfrag{016}[l][l]{SAO}
\psfrag{023}[c][c]{Energy [J]}
\psfrag{024}[r][r]{VP9 }
\psfrag{027}[r][r]{HEVC }
\psfrag{026}[r][r]{H.264 }
\psfrag{025}[r][r]{H.263 }
\psfrag{007}[r][r]{}
\psfrag{010}[r][r]{}
\psfrag{013}[r][r]{}
\psfrag{015}[r][r]{$E_\mathrm{dec}$}
\psfrag{014}[r][r]{$\hat E$}
\psfrag{012}[r][r]{$E_\mathrm{dec}$}
\psfrag{011}[r][r]{$\hat E$}
\psfrag{009}[r][r]{$E_\mathrm{dec}$}
\psfrag{008}[r][r]{$\hat E$}
\psfrag{006}[r][r]{$E_\mathrm{dec}$}
\psfrag{005}[r][r]{$\hat E$}
\includegraphics[width=.48\textwidth]{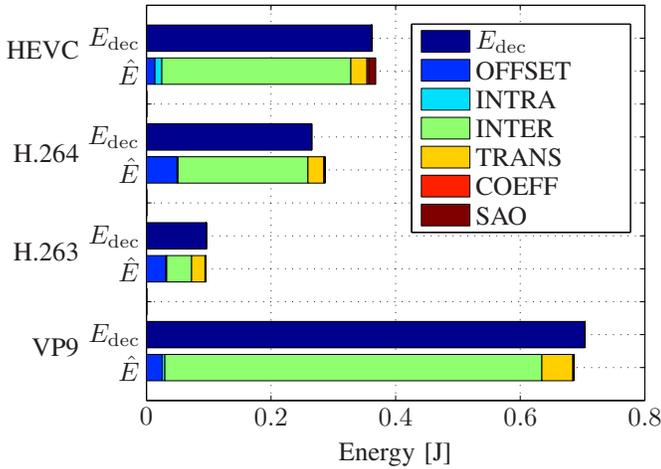}
\vspace{-.5cm}
\caption{Measured and estimated decoding energies (FFmpeg) for sequence Foreman coded at a nearly constant objective quality (PSNR $\approx 34.3$dB). The upper bars correspond to the measured energy $E_\mathrm{dec}$, the lower, stacked bars to the estimated energy $\hat E$. Each segment of the stacked bars represents the accumulated energies of the categories presented in Section \ref{sec:features}.   }
\label{fig:result_bars}
\end{figure}

To visualize how the results of our investigations can be interpreted we plot measured and estimated energies for a showcase sequence in Figure \ref{fig:result_bars}. 
We chose the QPs such that the bit streams are coded at an approximately constant PSNR in all codecs. Details are given in Table \ref{tab:foreman}. 

\begin{table}[t]
\renewcommand{\arraystretch}{1.3}
\caption{Objective visual quality, bit stream file size, and QP of the Foreman sequence coded using the different codecs. }
\label{tab:foreman}
\vspace{-.5cm}
\begin{center}
\begin{tabular}{l|c|c|c|c}
\hline
Codec & HEVC & H.264 & H.263 & VP9 \\
 \hline
 YUV-PSNR & $34.304$dB & $34.250$dB & $34.272$dB & $34.283$dB \\
File size & $20.3$kB & $31.3$kB & $103$kB & $28.7$kB\\  
QP & $33$ & $32$ & $8$ & $42$\\ 
 \hline
\end{tabular}
\end{center}
\end{table}

In Figure \ref{fig:result_bars}, the dark blue bars correspond to the measured energy $E_\mathrm{dec}$. Below, the stacked bars represent the estimated energy $\hat E$ where we can see that for each codec, it is very close to the measured energy. Each element of the stacked bar corresponds to a feature category as defined in Table \ref{tab:featList} where we can see that most energy is used for inter prediction. 

Furthermore, we can see that H.263-decoding, due to a very low complexity, consumes least energy. In contrast, the decoding energy of VP9 is relatively high which was not expected. Analyzing the distribution of the estimated energy on the features indicates that the fractional pel filtering process (feature {\footnotesize{frac}}) consumes a relatively large amount of energy in comparison to the other codecs.  



\section{Conclusions}
\label{sec:concl}
In this paper we have shown that a feature based approach to model the energy consumption of state-of-the-art video decoders is highly suitable to accurately estimate the decoding energy. Estimation errors are found to be lower than $8\%$ for all tested decoder implementations. In future work, the proposed models can be used to encode decoding energy saving bit streams.

\section*{ACKNOWLEDGEMENT}
This work was financially supported by the Research Training Group 1773 ``Heterogeneous Image Systems'', funded by the German Research Foundation (DFG).

\bibliographystyle{IEEEtran}
\bibliography{IEEEabrv,literatureNeu}

\end{document}